\documentclass[reprint,amsmath,amssymb,aps,
]{revtex4-1}

\usepackage[T1]{fontenc}
\usepackage{nicefrac}
\usepackage{hyperref}
\usepackage{amsmath}
\usepackage{amssymb}
\usepackage{graphicx}
\usepackage{color}
\usepackage{ulem}

\newcommand{\psec}[1]{\emph{#1.---}}
\newcommand{\B}[1]{\textcolor{black}{#1}} 
 
\newcommand{\lucas}[1]{\textcolor{black}{#1}}

\begin{document}

\title{Easing cosmic tensions with an open and hotter universe} 

\author{Benjamin Bose}
\affiliation{D\'{e}partement de Physique Th\'{e}orique, Universit\'{e} de Gen\`{e}ve, \\ 24 quai Ernest Ansermet, 1211 Gen\`{e}ve 4, Switzerland}
\author{Lucas Lombriser}
\affiliation{D\'{e}partement de Physique Th\'{e}orique, Universit\'{e} de Gen\`{e}ve, \\ 24 quai Ernest Ansermet, 1211 Gen\`{e}ve 4, Switzerland}

\date{\today}

\begin{abstract}
    Despite the great observational success of the standard cosmological model some discrepancies in the inferred parameter constraints have manifested among a number of cosmological data sets.
    These include a tension between the expansion rate of our Cosmos as inferred from the cosmic microwave background (CMB) and as found from local measurements, the preference for an enhanced amplitude of CMB lensing, a somewhat low quadrupole moment of the CMB fluctuations as well as a preference for a lower amplitude of matter fluctuations in large-scale structure surveys than inferred from the CMB.
    We analyse these observational tensions under the addition of spatial curvature and a free CMB background temperature that may deviate from its locally measured value.
    With inclusion of these parameters, we observe a trend in the parameter constraints from CMB and baryon acoustic oscillation data towards an open and hotter universe with larger current expansion rate, standard CMB lensing amplitudes, lower amplitude of matter fluctuations, and \lucas{marginally} lower CMB quadrupole moment, consistently reducing the individual tensions among the cosmological data sets.
    Combining this data with local distance measurements, we find a preference for an open and hotter universe beyond the 99.7\% confidence level.
    Finally, we briefly discuss a local void as a possible source for a deviation of the locally measured CMB temperature from its background value and as mimic of spatial curvature for CMB photons. \lucas{This interpretation implies a $\sim$20\% underdensity in our local neighbourhood of $\sim$10--100~Mpc in diameter, which is well within cosmic variance}.
\end{abstract}

\maketitle

\normalem

%%%%% INTRODUCTION %%%%%
\psec{Introduction}
%%%%%%%%%%%%%%%%%%%%%%%%
%
The standard model of cosmology ($\Lambda$CDM) has been very successful in reproducing the wealth of cosmological observations conducted over the past few decades.
Despite its successes, there remain a number of smaller and larger tensions among the data sets.
Perhaps the oldest yet relatively small discrepancy ($<2\sigma$) is the measurement of a somewhat low quadrupole moment in the cosmic microwave background (CMB) with respect to its $\Lambda$CDM prediction.
This was already noticed with COBE~\cite{1996ApJ...464L..17H} and has motivated studies of alternative cosmologies in the three decades since.
Another discrepancy in the CMB ($\lesssim3\sigma$) is observed in the imprint of the effect of weak gravitational lensing on the CMB spectra, which is enhanced with respect to the lensing that would be caused by the cosmological parameters inferred from early-time CMB physics~\cite{Aghanim:2018eyx}.
When considering the power spectrum of the reconstructed lensing potential, however, the preference for this enhancement is reduced again, which may perhaps be another indicator of an underlying inconsistency.
Clearly the strongest tension is observed in measurements of the current expansion rate of the Cosmos, where specifically CMB data and the local distance ladder measurement are in $4.4\sigma$ disagreement~\cite{Riess:2019cxk}.
Finally, another discrepancy ($\sim 2\sigma)$ manifests between the current amplitude of large-scale matter density fluctuations as predicted by the cosmological parameters inferred from the CMB and measurements from large-scale structure surveys~\cite{Asgari:2020wuj,Abbott:2017wau}.

Many exotic explanations and new physics models have been invoked to remedy these tensions 
\lucas{(see Ref.~\cite{DiValentino:2021izs} for a recent review)}.
But also the variation of more conventional parameters has been considered.
Specifically, the impact on parameter constraints and cosmic tensions from nonvanishing spatial curvature~\cite{Handley:2019tkm,DiValentino:2019qzk} or the variation of the CMB temperature~\cite{Ade:2015xua,Yoo:2019dyl,Ivanov:2020mfr,Bengaly:2020vly} have separately been studied.
While the variation of these parameters can reduce discrepancies among part of the data, the preference for new parameters is typically reduced when complementing the analysis with other data sets.
For instance, preferences in the CMB data for spatial curvature, varied CMB temperature, or a modified CMB lensing amplitude are suppressed when including baryon acoustic oscillation data (BAO) or the power spectrum of the reconstructed CMB lensing potential~\cite{Aghanim:2018eyx}.
As we will show in this \emph{Letter}, it is crucial to vary both spatial curvature and the CMB background temperature simultaneously to consistently ease the tensions among all the data sets.

We will first discuss the effects on the CMB power spectra and cross correlations from varying spatial curvature and the background temperature as well as the implications of that for the cosmic tensions.
We will then perform a parameter estimation analysis with different combinations of CMB, CMB lensing, BAO, and background expansion data and compare the constraints inferred on the extended parameter space against bounds from large-scale structure surveys and the local distance ladder.
Finally, we will discuss a local void as a possible source for a change of the locally measured CMB temperature with respect to that of the background and apparent spatial curvature.

%%%%%%% TENSIONS %%%%%%%
\psec{An open and hotter universe}
%%%%%%%%%%%%%%%%%%%%%%%%
%
To understand the impact on CMB fluctuations from nonvanishing spatial curvature and the change of the current CMB background temperature $T_0$ from the local FIRAS measurement $T_{\rm FIRAS} = (2.72548\pm0.00057)$~K~\cite{Fixsen:1996nj}, let us first examine how the early-time effects on the temperature and polarisation anisotropy power spectra and cross correlations transform under these variations.
Since recombination physics is governed by the ratios between the baryonic, cold dark matter, and photon energy densities, it is not difficult to deduce a scaling of cosmological parameters that keeps the oscillations in the CMB anisotropies fixed.
To preserve these ratios at the given recombination temperature $T_*$, using $T_*=T_0(1+z_*)$, we can simply fix $H_0^2 \Omega_i/T_0^3$, $i=m,b$, where $z_*$ is the redshift of recombination, $H_0$ is the Hubble constant, and $\Omega_b$ and $\Omega_m$ denote the fractional baryonic and total matter energy density parameters.
At a closer inspection, the situation is more complicated, but this simple relation holds true even for a more detailed analysis, as recently conducted by Ref.~\cite{Ivanov:2020mfr}.
Furthermore, the CMB acoustic peak positions are determined by $\ell_A = \pi D_A/s_*$, where $D_A$ is the angular diameter distance to recombination and $s_*$ is the sound horizon at recombination.
To preserve these positions under variation of $T_0$, we can therefore either change $H_0$ or we change the spatial curvature, giving rise to a geometric degeneracy between these three parameters.
Finally, a variation of $T_0$ also changes the overall amplitude of the anisotropies, which can be absorbed into a rescaling of the initial amplitude of density fluctuations $A_s$.

Of course, these degeneracies only apply to the primary anisotropies.
Given the recombination temperature, the measurement of the current CMB temperature $T_0$ is essentially a measurement of how much the universe has expanded since recombination.
Hence, a change in $T_0$ will therefore alter the late-time integrated Sachs-Wolve (ISW) effect and CMB lensing.
Whereas the detectability of changes in the ISW effect are strongly limited by cosmic variance, our observations are very sensitive to changes in the effect of CMB lensing.
It is also worth noting that varying the amount of expansion since recombination as well as the amplitude of initial perturbations $A_s$ will change the amplitude of current matter density fluctuations $\sigma_8$.

For illustration, let us for example consider a rise in the temperature $T_0$ from the FIRAS value.
To keep the primary CMB anisotropies fixed, we will need to increase $H_0^2\Omega_m$ and $H_0^2\Omega_b$.
We could now lower $H_0$ to keep the peak positions fixed or we could introduce negative spatial curvature $k<0$ to keep $H_0$ fixed or even raise it to bring it into accordance with local measurements.
Finally, we will also need to raise $A_s$ to keep the amplitude of the temperature fluctuations $\delta T_0/T_0$ fixed.
In contrast, the increase of $T_0$ changes the secondary anisotropies.
By raising $\Omega_m$, we also lower the ISW effect and reduce the small discrepancy between prediction and measurement of the quadrupole moment.
Another interesting effect is that because we raise $A_s$ and $T_0$ we no longer need to enhance the lensing amplitude $A_L$ from its standard value of unity.
Finally, since higher $T_0$ implies less expansion since recombination until today, this also lowers $\sigma_8$, despite the increase of $A_s$.

%%%%%%% RESULTS %%%%%%%%
\psec{Observational results}
%%%%%%%%%%%%%%%%%%%%%%%%
%
Given the promising phenomenological features arising from introducing variations in spatial curvature and $T_0$, let us now turn to a data analysis to explore their direct impact on the measurements.
As the base data sets in this analysis we will use the CMB temperature, polarisation, and cross correlation data from Planck~\footnote{We use Planck 2015 data~\cite{Ade:2015xua}. The implementation of the Planck 2018 likelihood into {\sc cosmosis}~\cite{Zuntz:2014csq} is currently
not available,
%undergoing testing,
and we defer an analysis with the newer data to a later date, although we do not expect significant changes in our findings. Also note that we adopt the default vanishing total neutrino mass of the code.} (TT, TE, EE + lowE).
In addition to that, we will also perform a further analysis with the additional inclusion of the CMB lensing power spectrum.
The reason for the separate analyses being that we want to study changes in the constraints on $A_L$ when including the reconsructed lensing data and when omitting it.
We then perform a set of analyses, where in addition to the CMB measurements, we use the BAO observations from the Baryon Oscillation Spectroscopic Survey (BOSS) DR12 release~\cite{Alam:2016hwk}.
Finally, we compare the results from the CMB and CMB+BAO analyses against local $H_0$ measurements from Ref.~\cite{Riess:2019cxk} (R19). We also conduct analyses using all data sets.

For the parameter estimations on this data we perform Markov Chain Monte Carlo (MCMC) sampling using the {\sc cosmosis}~\cite{Zuntz:2014csq} package.
As our base cosmological parameters we choose
\begin{equation}
\theta_{\rm base} = \{\Omega_b, \Omega_m, H_0, n_s, \tau, A_s \} \,.
\end{equation}
We first conduct an analysis for the base parameters with the different combinations of the data sets.
We then augment the parameter space with the additional parameters: the fractional energy density parameter attributed to spatial curvature $\Omega_k$, the modification of the lensing amplitude $A_L$, and the free CMB background temperature $T_0$.
The parameters are then constrained for the different combinations of the data as well as the alternate combinations in the addition of the extra parameters.

%%% FIG %%%
\begin{figure}
\centering
 \resizebox{0.475\textwidth}{!}{
 \includegraphics{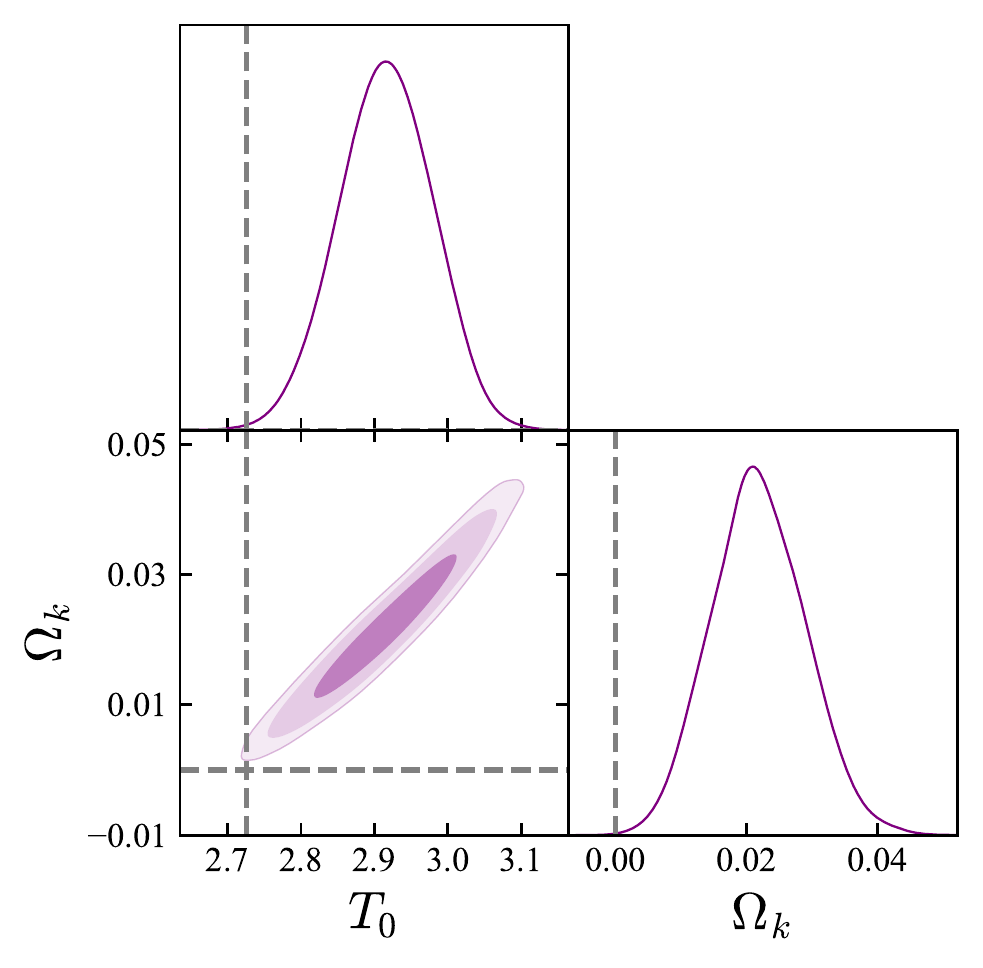}
 }
\caption{The combination of all data sets (Planck with lensing, BAO, and R19) shows a preference for an open ($\Omega_k>0$) and hotter ($T_{\rm 0} > T_{\rm FIRAS} = 2.7255$) universe. The contours indicate 68.3\%, 95.4\%, and 99.7\% confidence levels.
}
\label{fig:oktcmb}
\end{figure}
%%% FIG %%%

Our main result is given in Fig.~\ref{fig:oktcmb}, which shows a preference for an open ($\Omega_k>0)$ and hotter ($T_0>T_{\rm FIRAS}$) universe  beyond the 99.7\% confidence level from the combination of all the data.  We have performed two analyses, one sampling $\theta_{\rm base}$ with $A_s$ and another with $\sigma_8$ instead. The posteriors do not shift significantly in the $T_0$ vs $\Omega_k$ plane, but we have noted a shift of the mean of $\sigma_8$ to slightly larger values when sampling in $A_s$ ($\sigma_8$ is calculated as a derived parameter in this case). We leave this curiosity to a future work where we consider other large-scale structure data sets such as the KiDS cosmic shear measurements \cite{Asgari:2020wuj}. Fig.~\ref{fig:oktcmb} shows results for the sampling in $A_s$.

%%% FIG %%%
\begin{figure}
\centering
 \resizebox{0.45\textwidth}{!}{
 \includegraphics{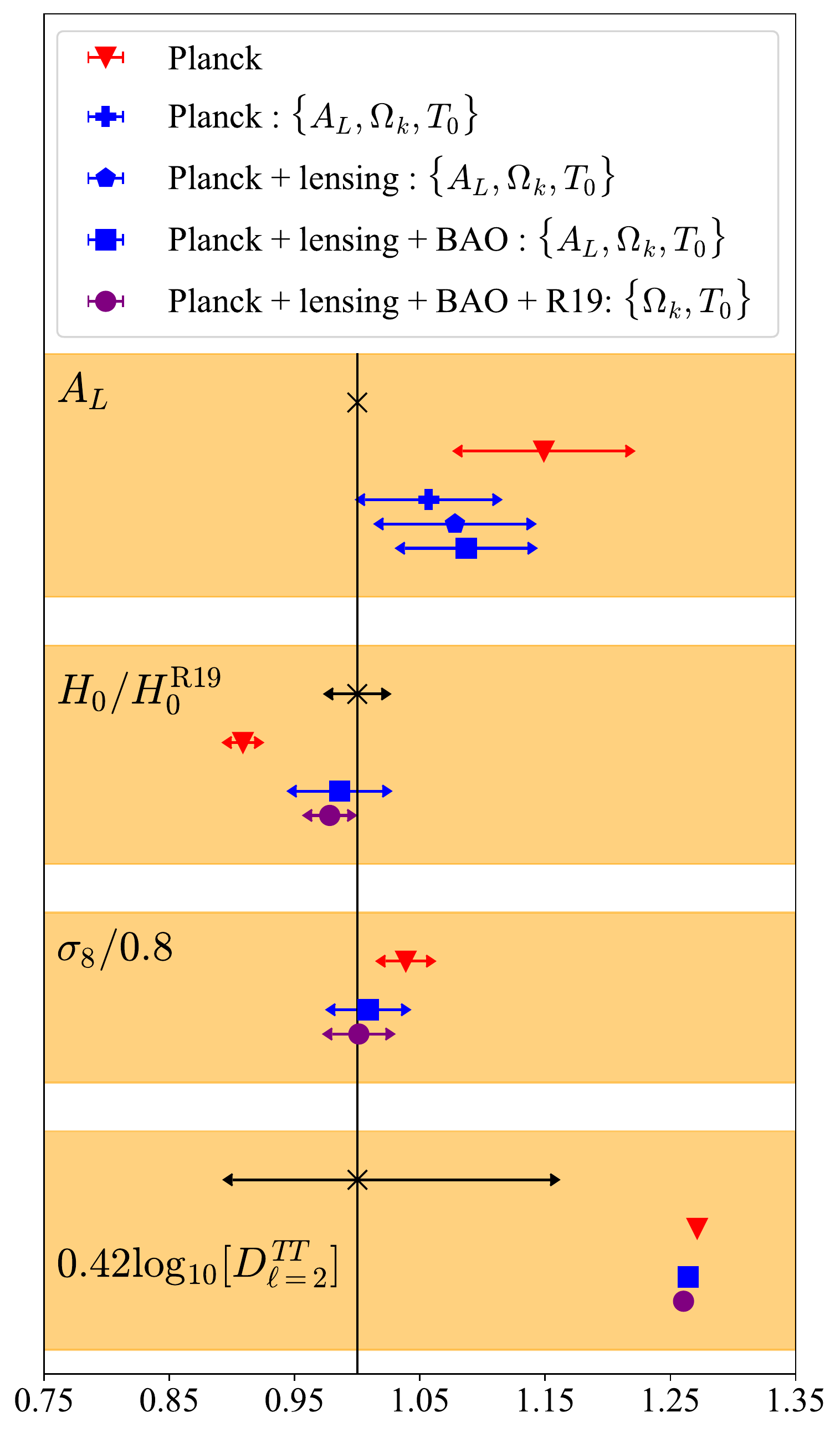}
 }
\caption{With the additional variation of $\Omega_k$ and $T_{\rm 0}$ there is no longer any noteworthy preference for an enhanced amplitude of CMB lensing $A_L>1$.
The CMB and BAO data also favour a larger current expansion rate $H_0$ in better agreement with local measurements as well as a lower amplitude of matter fluctuations $\sigma_8$ as preferred by large-scale structure surveys.
A further effect is a \lucas{marginal} lowering of the predicted CMB quadrupole towards the measurement. Error bars represent marginalized 68.3\% confidence levels.
}
\label{fig:parshift}
\end{figure}
%%% FIG %%%

In Fig.~\ref{fig:parshift}, we summarise how the parameter constraints shift under inclusion of $A_L$, $\Omega_k$, and $T_0$ as well as additional data.
We find that with the shift to negative spatial curvature ($k<0$, $\Omega_k>0$) and a hotter temperature ($T_{\rm 0}>T_{\rm FIRAS}$) there is no longer any noteworthy preference for an enhanced amplitude of CMB lensing $A_L>1$.
This is regardless of whether the reconstructed power spectrum of the CMB lensing potential or BAO data is included or not. The CMB and BAO data also favour a larger current expansion rate $H_0$, in better agreement with local measurements as well as a lower amplitude of matter fluctuations $\sigma_8$ as preferred by large-scale structure surveys.
In particular KiDS has reported a measurement of $S_8\equiv\sigma_8 \sqrt{\Omega_m/0.3}=0.759^{+0.024}_{-0.021}$~\cite{Asgari:2020wuj}.
 Note, however, that this constraint does not include a variation of spatial curvature and can therefore not be directly compared to the results in Fig.~\ref{fig:parshift}. %
\B{A recent joint KiDS-2dFLens-BOSS analysis \cite{Troster:2020kai} including spatial curvature  has reported $\Omega_k = 0.011^{+0.054}_{-0.057}$. 
They also find a positive degeneracy between $\Omega_k$ and $S_8$, consistent with our results.}
\B{A DES analysis~\cite{Abbott:2018xao} furthermore} reported a constraint of $\Omega_k = 0.16^{+0.09}_{-0.14}$ \lucas{and a supernovae analysis of Ref.~\cite{Yang:2020bpv} indicates a 2$\sigma$ preference for $\Omega_k>0$}, consistent with the trend towards an open universe exhibited in Fig.~\ref{fig:oktcmb}.
Finally, a further effect of the open and hotter universe illustrated in Fig.~\ref{fig:parshift} is a \B{marginal} lowering of the CMB quadrupole moment towards the measurement.

%%%%%%%%% VOID %%%%%%%%%
\psec{Local void as a possible interpretation}
%%%%%%%%%%%%%%%%%%%%%%%%
%
As a possible candidate for the enhanced CMB background temperature over its locally measured value, let us briefly examine the effects that would arise from residing in a local void.
As a simple picture of the local underdensity, we may consider a top-hat matter density fluctuation.
This implies that our local universe can be interpreted within the separate universe ansatz.
The local expansion is then described by the usual Friedmann equations, which are however governed by the local matter density rather than that of the background.
The local and background metrics are related by a time $t$ dependent transformation $\hat{a}(t)=C(t)a(t)$, where $\hat{a}$ and $a$ denote the local and background scale factors and $C=(1+\delta)^{-1/3}$ is specified by the matter density fluctuation $\delta$~\cite{Lombriser:2019ahl}.
\lucas{Importantly, number conservation of the black-body photon gas for its changing volume implies $\hat{T}_0=T_0/C_0$.}
Thus, in an \lucas{evolving} local underdensity $\delta_0<0$ the CMB temperature is lower and the expansion rate is higher than in the background.
\lucas{In contrast, for photons entering a virialised structure, their temperature is only affected by the local potential well.}
\lucas{One now has two options of interpreting the CMB data.
Either one corrects the monopole temperature, just as one removes the dipole before inferring the cosmological parameters.
Or, less naturally, one interprets the data by associating the local temperature to the cosmological background, which leads to an effective cosmology.
In this case, however, one needs to perform the same interpretation for the distance ladder, which amounts to a rescaling of the distance anchor to a corresponding effective distance in the background.}

\lucas{Here, we adopt the first approach.
Ref.~\cite{Lombriser:2019ahl} adopted the second.
The two approaches are related through a conformal metric transformation with $C(t)$, and thus a change of frame~\cite{Lombriser:2019ahl}.
Note that independently of the void, a local conformal factor may also be motivated simply as an available geometric freedom that must be accounted for and observationally constrained~\cite{Visser:2015iua}.
In Ref.~\cite{Lombriser:2019ahl} the rescaled expansion rate caused by the correction of the distance anchor was approximated by the local expansion rate.
More accurately it is simply $C_0H_0$, which implies that the local void can be less underdense than inferred in Ref.~\cite{Lombriser:2019ahl} and statistically even more likely given cosmic variance.
To be clear, in this picture, the $H_0$ value measured with the distance ladder should be considered the correct current expansion rate of the cosmological background.
However, in either approach the CMB and local distance ladder measurements of $H_0$ agree as long as one does not mix the two approaches.
}
\lucas{Finally, it is also worth noting that}
the angular diameter distance transforms as $\hat{D}_A = C_0 D_A$ and gets elongated.
\lucas{To preserve the measured CMB angular peak positions, which remain unaffected by the conformal transformation, one can thus change $H_0$ or $\Omega_k$.
Hence, omitting the effect of the local void can mimic spatial curvature, although with different magnitudes for data at different distances.
}

\B{
To give a more quantitative picture, we can adopt the mean value of $T_0 = 2.95$~K obtained from our Planck analysis with lensing and BAO included (blue squares in Fig.~\ref{fig:parshift}) to infer a local void of $\delta_0 = -0.21$.
The size of the void \lucas{cannot be inferred from our data}, and we leave its computation and statistical significance for future work.
It is worth pointing out, however, that Ref.~\cite{Lombriser:2019ahl} found a void of size \lucas{$\sim$10--100}~Mpc \lucas{in diameter} to be sufficient to reconcile the Hubble constants of R19 and Planck,
%Given the underdensity we estimate, one could consider larger voids ($\geq 100$ Mpc), consistent with other local  measurements \cite{Boehringer:2019xmx}.
and measurements of galaxy groups~\cite{Karachentsev:2018ysz} and clusters~\cite{Boehringer:2019xmx} \lucas{as well as peculiar velocities~\cite{Tully:2019ngb}} in such a neighborhood indeed show a preference for a local underdensity.}
%~\cite{Karachentsev:2018ysz,Boehringer:2019xmx}.}
%\BB{Further, using simulations, it was shown in Ref.~\cite{Wojtak:2013gda} that a void of  this size would allow for a $4\%$ shift in the background $H_0$ at $1\sigma$ confidence.}
%Note that we do not include the R19 measurement of $H_0$ in our underdensity estimate as it would have an effect on the measured value of $H_0$ of $\sim 4\%$.
%Interestingly this would move $H_0$ in the right direction in Fig.~\ref{fig:parshift}.
%
\lucas{Analysing the supernovea data used in the distance laddder, Refs.~\cite{Kenworthy:2019qwq,Wu:2017fpr} found an upper bound on the radius of the local void of $R\lesssim100$~Mpc.}

\lucas{
%%%%%%% MORE DATA %%%%%%
\psec{Further comparisons}
%%%%%%%%%%%%%%%%%%%%%%%%
%
}
\lucas{Our Planck analysis with lensing, BAO, and fixed $A_L=1$ yields $H_0=70.05 \pm 1.73$~km/s/Mpc for free $T_0$ and $\Omega_k$. This is consistent with R19 at about the $2\sigma$-level such that the inclusion of the constraint in the analysis is justified.}
\B{One may also consider the inclusion of further data sets such as Lyman-$\alpha$ constraints, which when combined with BAO measurements, prefer a low value of $H_0$.
Ref.~\cite{Cuceu:2019for} finds $H_0 = 68.1 \pm 1.1$~km/s/Mpc.}
\lucas{In principle, this is consistent with our constraint, and moreover, \B{Ref.~\cite{Cuceu:2019for} does not vary spatial curvature, which if done would increase error bars and the consistency among the two values.}}
\lucas{We stress however the importance of carefully reassessing the assumptions made in a measurement before comparing the inferred parameter values and combining data sets.
In the case of the Lyman-$\alpha$ constraint of Ref.~\cite{Cuceu:2019for}, Big Bang Nucleosynthesis (BBN) constraints on $\Omega_bh^2$ are used to infer the sound horizon at the baryon drag epoch.
For consistency, these must be rescaled to account for the change of $T_0$ from $T_{\rm FIRAS}$, just as it equivalently rescales $\Omega_bh^2$ inferred from the CMB oscillations.
%Hence, for consistency with the $H_0$ value we infer here from the CMB with free $T_0$, this constraint should be lifted.
}
%\BB{We note that this value is also consistent with R19 within $2\sigma$.}
%
\B{\lucas{Note that} temperatures at BBN do not change, but rather the extrapolation from the local temperature $\hat{T}_0$ differs from a homogeneous one when accounting for the local inhomogeneity.
}
%

%%%%%% CONCLUSIONS %%%%%
\psec{Conclusions}
%%%%%%%%%%%%%%%%%%%%%%%%
%
Concordance cosmology has been very successful in reproducing our cosmological observations, but some smaller and larger discrepancies in the inferred parameter constraints have manifested among a number of our data sets.
We analysed these observational tensions under the addition of spatial curvature and a free CMB background temperature allowed to deviate from its locally measured value.
The inclusion of these parameters produce a trend in the constraints inferred from CMB and BAO data towards an open and hotter universe with larger current expansion rate, standard CMB lensing amplitudes, lower amplitude of matter fluctuations, and \lucas{marginally} lower CMB quadrupole moment.
This trend consistently eases the individual tensions among the cosmological data sets.
As a consequence, when combining the data with local distance measurements, we found a preference for an open and hotter universe beyond the 99.7\% confidence level.
We then briefly discussed how a local void may serve as an explanation for the deviation of the CMB background temperature from its locally measured value and mimic spatial curvature for CMB photons.
\lucas{This interpretation implies a $\sim$20\% underdensity in our local neighbourhood of $\sim$10--100~Mpc in diameter, which is well within cosmic variance.}

%%%% ACKNOWLEDGMENTS %%%
%
The authors thank Charles Dalang for useful discussions. This work was supported by a Swiss National Science Foundation Professorship grant (No.~170547).
Please contact the authors for access to research materials.

%%%%%% BIBLIOGRAHY %%%%%%
\bibliographystyle{arxiv_physrev}
\bibliography{cosmic-tensions}
%%%%%%%%%%%%%%%%%%%%%%%%

\end{document}